\documentclass[twocolumn,showpacs,preprintnumbers,superscriptaddress,floatfix,nofootinbib]{revtex4}
\usepackage{amssymb}
\usepackage{amsmath}
\usepackage{graphicx}
\usepackage{dcolumn}
\usepackage{color}
\usepackage{bm}
\usepackage[subfigure]{graphfig}
\usepackage{makecell}

\begin{document}

\title{Photoproduction of $a_{2}(1320)$ in a Regge model}
\author{Xiao-Yun Wang}
\thanks{xywang@impcas.ac.cn}
\affiliation{Institute of Modern Physics, Chinese Academy of Sciences, Lanzhou 730000,
China}
\affiliation{University of Chinese Academy of Sciences, Beijing 100049, China}
\affiliation{Research Center for Hadron and CSR Physics, Institute of Modern Physics of
CAS and Lanzhou University, Lanzhou 730000, China}
\author{Alexey Guskov}
\thanks{avg@jinr.ru}
\affiliation{Joint Institute for Nuclear Research, Dubna 141980, Russia}

\begin{abstract}
In this work, the photoproduction of $a_{2}(1320)$ off a proton target is
investigated within an effective Lagrangian approach and the Regge model.
The theoretical result indicates that the shapes of {the} total and
differential cross sections of the $\gamma p\rightarrow a_{2}^{+}n$
reaction within the Feynman (isobar) model are much different from that of
the Reggeized treatment. The obtained cross section is compared with the
existing experimental results at low energies. The $a_{2}(1320)$ production
cross section at high energies can be tested by the COMPASS experiment,
which can provide important information for clarifying the role of the
Reggeized treatment at that energy range.
\end{abstract}

\maketitle

\section{Introduction}

Within the past decades great progress has been achieved in hadron
spectroscopy \cite{mn10,liu14}. Especially, inspired by the observation of
exotic states \cite{mn10,liu14} (such as the candidate for the tetraquark
\cite{ra14} or pentaquark \cite{ra15} state etc.), the underlying structure
of these states attracts much attention both in theory and experiment.
Observation of the exclusive photoproduction of exotic hadronic states off baryons, proposed in the Refs. \cite{liu8, he09, lin14, liu15, karliner16}, is the most direct way to get
information about their nature. At higher energies such
processes can be described in terms of Regge trajectory exchanges \cite%
{pd77,vg09}. In Ref. \cite{wt03}, the $\eta $ and $\eta'$
photoproduction were studied with the Reggeized model. It is found that the
Reggeized model gives a good description for these reactions in and beyond
the resonance region. Thus the photoproduction reaction at high energy may be appropriate to
study the role of Reggeized treatment.

In the 1950s, Regge proved the importance of extending the angular momentum $%
J$ to the complex field \cite{regge59,regge60}. For more general reviews
about the Regge theory, see Refs. \cite{rj71,tc06,jk87}. Later, the
exchange of dominant meson Regge trajectories {was} used to successfully
describe the hadron photoproduction \cite{mg97,gg11,he14,ew14}. However, {%
there is one question}: has the Regge trajectory approach been well tested
by experiment at higher energies?

In the past, the $a_{2}(1320)$ ($\equiv a_{2}$) photoproduction was
extensively studied. The estimation of the exclusive photoproduction cross section for $a_2$ has been performed according to the one-pion exchange (OPE) mechanism with absorption in \cite{Hog66}. The experimental results for the values and energy dependence of this cross section at relatively low ($<$20 GeV) energies are quite consistent with the prediction of this model \cite{gt93}. Nevertheless the energy range covered by the existing experimental data is not enough to distinguish between the OPE prediction and  the Regge trajectory approach \cite{mo66}.

The COMPASS experiment at CERN, uses the muon beam, can significantly enlarge the available
energy range of virtual photons up to about 150 GeV. COMPASS has a good opportunity to contribute to the study of exotic charmonia via their photoproduction. However the uncertainties of the theoretical description of photon-nucleon interaction at high energies complicate this task. The process of $\gamma ^{\ast }p\rightarrow a_{2}^{+}n$, where $\gamma ^{\ast }$ is a virtual photon, has quite good experimental signature and can be used as a benchmark. Possibility to use $a_2$ photoproduction as a benchmark for study of exotic hadrons is discussed also in \cite{cz1}.
It is also significant to carry out more
theoretical studies on the $\gamma p\rightarrow a_{2}^{+}n$ process in order
to clarify the role of the Reggeized treatment.

Moreover, due to the vector meson dominance (VMD) assumption, a photon can
interact with a vector meson, which means that the $\gamma p\rightarrow
a_{2}^{+}n$ reaction can also{\ proceed through} the vector meson dominance
(VMD) mechanism \cite{tb65,tb69,th78}. Thus {the $a_{2}$ photoproduction
mechanism} is also an interesting issue.

In this work, the $\gamma p\rightarrow a_{2}^{+}n$ reaction is investigated
using an effective Lagrangian approach and the Regge model. In addition to
the $\pi $ exchange, the contributions from {the} VMD mechanism is also
considered. The differential cross section of the $\gamma p\rightarrow
a_{2}^{+}n$ reaction is also calculated, which could be tested by further
COMPASS experiment.

This paper is organized as follows. After the introduction, we present the
formalism and the main ingredients which are used in our calculation. The
numerical results and discussions are given in Sec. III. In Sec. IV, we give
a detailed illustration of the possibility of the experimental test at
COMPASS. Finally, the paper ends with a brief summary.

\section{Formalism}

In the present work, an effective Lagrangian approach in terms of hadrons is
adopted, which is an important theoretical method in investigating various
scattering processes \cite{zou03,xyw15,xy,epl15,epja15,xy2940}.

Figure 1 describes the basic tree-level Feynman diagrams for the $%
a_{2}(1320)$ photoproduction process through general $\pi $ exchange [Fig.
1(a)] and vector meson dominance (VMD) mechanism [Fig. 1(b)] \cite%
{tb65,tb69,th78}.

\begin{figure}[tbph]
\begin{center}
\includegraphics[scale=0.5]{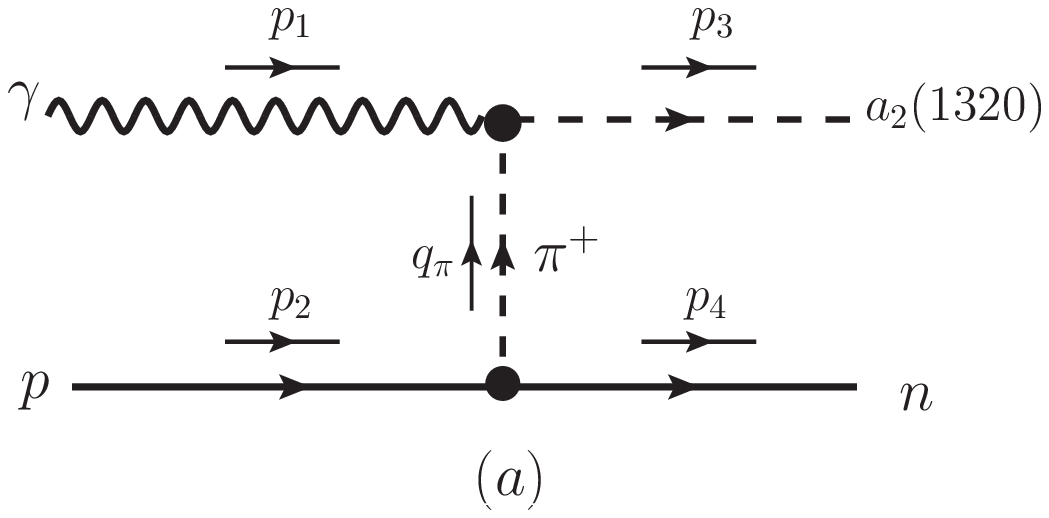} \includegraphics[scale=0.5]{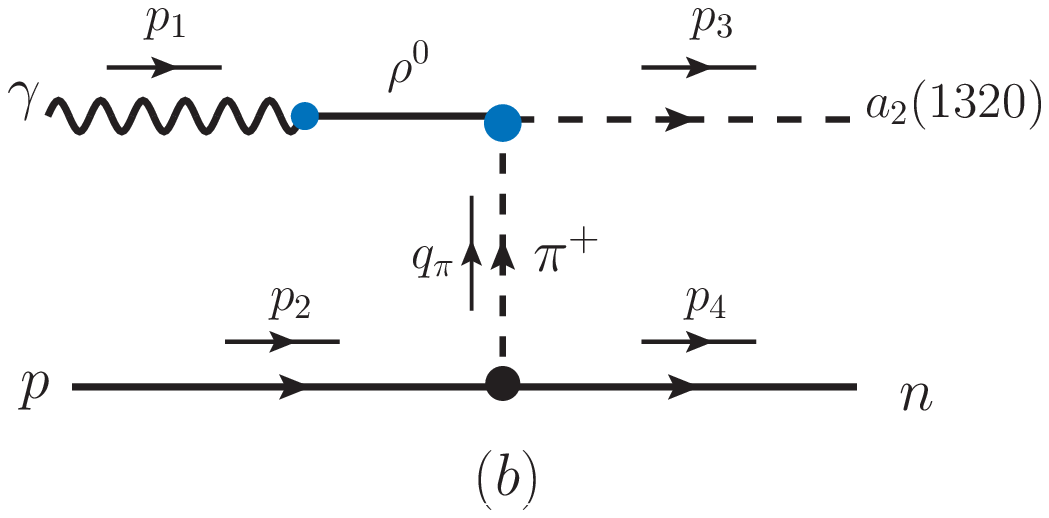}
\end{center}
\caption{(a): Feynman diagrams for the $\protect\gamma p\rightarrow a_{2}n$
reaction via $\protect\pi $ exchange. (b) {same as in} (a), but for the case
in the frame of{\ the} VMD model.}
\end{figure}

To gauge the contribution of this diagram, we need to know the relevant
effective Lagrangian densities.

For the {$\pi NN$ interaction vertex} we take the effective pseudoscalar
coupling \cite{kt98},%
\begin{equation}
\mathcal{L}_{\pi NN}=-ig_{\pi NN}\bar{N}\gamma _{5}\vec{\tau}\cdot \vec{\pi}N
\end{equation}%
where $\vec{\tau}$ is the Pauli matrix, while $N$ and $\pi $ stand for the
fields of the nucleon and the pion, respectively. The coupling constant
of the $\pi NN$ interaction was given in many theoretical works, and we take
$g_{\pi NN}^{2}/4\pi =12.96$ \cite{lzw,vb11}.

The commonly employed Lagrangian densities for $a_{2}\pi \gamma $ {and} $%
a_{2}\pi \rho $ couplings are \cite{nl76,jb76,kb80,ze84}:
\begin{eqnarray}
\mathcal{L}_{a_{2}\pi \gamma } &=&\frac{g_{a_{2}\pi \gamma }}{m_{\pi }^{2}}%
\epsilon _{\mu \nu \alpha \beta }\partial ^{\mu }a_{2}^{\nu \sigma }\partial
^{\alpha }A^{\beta }\partial _{\sigma }\phi _{\pi }, \\
\mathcal{L}_{a_{2}\pi \rho } &=&\frac{g_{a_{2}\pi \rho }}{m_{\pi }^{2}}%
\epsilon _{\mu \nu \alpha \beta }\partial ^{\mu }a_{2}^{\nu \sigma }\partial
^{\alpha }\rho ^{\beta }\partial _{\sigma }\phi _{\pi },
\end{eqnarray}%
where $A^{\beta }$, $a_{2}^{\nu \sigma }$, $\rho ^{\beta }$ and $\phi _{\pi
} $ are the photon, $a_{2}$ meson, $\rho $ and $\pi $ fields. $m_{\pi }$ is
the mass of the $\pi $ meson. The coupling constant $g_{a_{2}\pi \gamma }$
and $g_{a_{2}\pi \rho }$ can be determined by the partial decay widths $%
\Gamma _{a_{2}\rightarrow \pi \gamma }$ and $\Gamma _{a_{2}\rightarrow \pi
\rho }$, respectively. With the above Lagrangian densities, {we} obtain
\begin{eqnarray}
\Gamma _{a_{2}\rightarrow \pi \gamma } &=&\frac{g_{a_{2}\pi \gamma }^{2}}{%
10\pi m_{\pi }^{4}}\left\vert \vec{p}_{\gamma }^{~\mathrm{c.m.}}\right\vert
^{5}, \\
\Gamma _{a_{2}\rightarrow \pi \rho } &=&\frac{g_{a_{2}\pi \rho }^{2}}{10\pi
m_{\pi }^{4}}\left\vert \vec{p}_{\rho }^{~\mathrm{c.m.}}\right\vert ^{5},
\end{eqnarray}%
with%
\begin{eqnarray}
|\vec{p}_{\gamma }^{~\mathrm{c.m.}}| &=&\frac{\lambda
^{1/2}(M_{a_{2}}^{2},m_{\pi }^{2},m_{\gamma }^{2})}{2M_{a_{2}}}, \\
|\vec{p}_{\rho }^{~\mathrm{c.m.}}| &=&\frac{\lambda
^{1/2}(M_{a_{2}}^{2},m_{\pi }^{2},m_{\rho }^{2})}{2M_{a_{2}}},
\end{eqnarray}%
where $\lambda $ is the K$\ddot{a}$llen function with $\lambda
(x,y,z)=(x-y-z)^{2}-4yz$. Using the partial decay widths of $a_{2}(1320)$ as
listed in {the} PDG book \cite{pdg}, {we} get $g_{_{a_{2}\pi \gamma
}}=0.539\times 10^{-2}$ GeV$^{-1}$ and $g_{_{a_{2}\pi \rho }}=0.268$ GeV$%
^{-1}$.

For the $a_{2}\gamma \pi $ interaction vertex we can also derive it
using the vector meson dominance (VMD) mechanism \cite{tb65,tb69,th78} on
the assumption that the coupling is due to a sum of intermediate vector
mesons. In the VMD mechanism for photoproduction, a real photon can
fluctuate into a virtual vector meson, which subsequently scatters from the
target proton. Under the VMD mechanism, the Lagrangian of depicting the
coupling of the intermediate vector meson $\rho $ with a photon is written
as
\begin{equation}
\mathcal{L}_{\rho \gamma }=-\frac{em_{\rho }^{2}}{f_{\rho }}V_{\mu }A^{\mu },
\end{equation}%
where $m_{\rho }^{2}$ and $f_{\rho }$ are the mass and the decay constant of
{the} $\rho $ meson, respectively. With the above equation, {we get} the
expression for the $\rho \rightarrow e^{+}e^{-}$ decay,%
\begin{equation}
\Gamma _{\rho \rightarrow e^{+}e^{-}}=\left( \frac{e}{f_{\rho }}\right) ^{2}%
\frac{8\alpha \left\vert \vec{p}_{e}^{~\mathrm{c.m.}}\right\vert ^{3}}{%
3m_{\rho }^{2}},
\end{equation}%
where $\vec{p}_{e}^{~\mathrm{c.m.}}$ indicate{s} the three-momentum of an
electron in the rest frame of the $\rho $ meson, while $\alpha =e^{2}/4\hbar
c=1/137$ is the electromagnetic fine structure constant. Thus, with the
partial decay width of $\rho \rightarrow e^{+}e^{-}$ \cite{pdg}%
\begin{equation}
\Gamma _{\rho \rightarrow e^{+}e^{-}}=7.04\text{ keV,}
\end{equation}%
we get the constant $e/f_{\rho }\simeq 0.06$.

To account for the internal structure of hadrons, we introduce
phenomenological form factors. For the vertex of $a_{2}\pi \rho $, the
following form factor is adopted \cite{xy,mosel98,mosel99},%
\begin{equation}
\mathcal{F}_{a_{2}\pi \rho }(q_{\pi }^{2})=\frac{m_{\rho }^{2}-m_{\pi }^{2}}{%
m_{\rho }^{2}-q_{\pi }^{2}}.
\end{equation}

For the vertices of $\pi NN$ and $a_{2}\pi \gamma $, three types of the form
factors are considered \cite{xy,mosel98,mosel99,vp16,fc16}: (i) the monopole
form factor%
\begin{equation}
\mathcal{F}_{\pi NN}(q_{\pi }^{2})=\mathcal{F}_{a_{2}\pi \gamma }(q_{\pi
}^{2})=\frac{\Lambda _{t}^{2}-m_{\pi }^{2}}{\Lambda _{t}^{2}-q_{\pi }^{2}},
\end{equation}%
(ii) the dipole form factor%
\begin{equation}
\mathcal{F}_{\pi NN}(q_{\pi }^{2})=\mathcal{F}_{a_{2}\pi \gamma }(q_{\pi
}^{2})=\left( \frac{\Lambda _{d}^{2}-m_{\pi }^{2}}{\Lambda _{d}^{2}-q_{\pi
}^{2}}\right) ^{2},
\end{equation}%
(iii) the exponential form factor%
\begin{equation}
\mathcal{F}_{\pi NN}(q_{\pi }^{2})=\mathcal{F}_{a_{2}\pi \gamma }(q_{\pi
}^{2})=\exp \left[ R^{2}\frac{\left( t-m_{\pi }^{2}\right) }{\left(
1-X_{L}\right) }\right],
\end{equation}
where $\Lambda _{t},\Lambda _{d}$ and $R$ are the free parameters, which can
be determined from the data in this work. $X_{L}$ is momentum fraction of
the proton carried by the neutron.

With the effective Lagrangian densities as listed above, the invariant
scattering amplitudes for the $\gamma (p_{1})p(p_{2})\rightarrow
a_{2}^{+}(p_{3})n(p_{4})$ process can be written as

\begin{eqnarray}
\mathcal{M}_{a} &=&\sqrt{2}\frac{g_{\pi NN}g_{a_{2}\pi \gamma }}{m_{\pi }^{2}%
}\frac{\mathcal{F}_{\pi NN}(q_{\pi }^{2})\mathcal{F}_{a_{2}\pi \gamma
}(q_{\pi }^{2})}{q_{\pi }^{2}-m_{\pi }^{2}}\bar{u}(p_{4})  \notag \\
&&\times \gamma _{5}\epsilon _{\mu \nu \alpha \beta }p_{3}^{\mu }T^{\nu
\sigma }(p3)p_{1}^{\alpha }\epsilon ^{\beta }(p_{1})(q_{\pi })_{\sigma
}u(p_{2})
\end{eqnarray}%
for Fig. 1(a), and%
\begin{eqnarray}
\mathcal{M}_{b} &=&\sqrt{2}\frac{g_{\pi NN}g_{a_{2}\pi \rho }}{m_{\pi }^{2}}%
\frac{e}{f_{\rho }}\frac{\mathcal{F}_{\pi NN}(q_{\pi }^{2})\mathcal{F}%
_{a_{2}\pi \rho }(q_{\pi }^{2})}{q_{\pi }^{2}-m_{\pi }^{2}}\bar{u}(p_{4})
\notag \\
&&\times \gamma _{5}\epsilon _{\mu \nu \alpha \beta }p_{3}^{\mu }T^{\nu
\sigma }(p3)p_{1}^{\alpha }\epsilon ^{\beta }(p_{1})(q_{\pi })_{\sigma
}u(p_{2})
\end{eqnarray}%
for Fig. 1(b). {Here} $\epsilon ^{\beta }(p_{1})$ and $T^{\nu \sigma }(p3)$
are the photon polarization vector and {the} polarization vector of the $%
a_{2}$, respectively{\ ,} $u(p_{2})$ and $\bar{u}(p_{4})$ are the Dirac
spinors for the initial proton and final {the} neutron, respectively.

To describe the behavior at high photon energy, we introduce the Regge
trajectories \cite{mg97,xy,ai05,tc07}%
\begin{equation}
\frac{1}{q_{\pi }^{2}-m_{\pi }^{2}}\rightarrow \mathcal{D}_{\pi }=(\frac{s}{%
s_{\text{scale}}})^{\alpha _{\pi }(t)}\frac{\pi \alpha _{\pi }^{\prime
}e^{-i\pi \alpha _{\pi }(t)}}{\Gamma \lbrack 1+\alpha _{\pi }(t)]\sin [\pi
\alpha _{K}(t)]},
\end{equation}%
where $\alpha _{K}^{\prime }$ is the slope of the trajectory and the scale
factor $s_{\text{scale}}$ is fixed at 1 GeV$^{2}$, while $%
s=(p_{1}+p_{2})^{2} $ and $t=(p_{2}-p_{4})^{2}$ are the Mandelstam
variables. In addition, the kaonic Regge trajectory $\alpha _{K}(t)$ {is}
\cite{mg97,xy,ai05,tc07}%
\begin{equation}
\alpha _{\pi }(t)=0.7(t-m_{\pi }^{2}).
\end{equation}

{With} $s=(p_{1}+p_{2})^{2}$, the unpolarized differential cross section for
the $\gamma (p_{1})p(p_{2})\rightarrow a_{2}^{+}(p_{3})n(p_{4})$ process at
the center of mass (c.m.) frame is given by
\begin{equation}
\frac{d\sigma }{d\cos \theta }=\frac{1}{32\pi s}\frac{\left\vert \vec{p}%
_{3}^{{~\mathrm{c.m.}}}\right\vert }{\left\vert \vec{p}_{1}^{{~\mathrm{c.m.}}%
}\right\vert }\left( \frac{1}{4}\sum\limits_{spins}\left\vert \mathcal{M}%
_{a/b}\right\vert ^{2}\right)
\end{equation}%
where $\theta $ denotes the angle of the outgoing $a_{2}^{+}$ meson relative
to {the} beam direction in the c.m. frame, while $\vec{p}_{1}^{{~\mathrm{c.m.%
}}}$ and $\vec{p}_{3}^{{~\mathrm{c.m.}}}$ are the three-momenta of {the}
initial photon beam and {the} final $a_{2}^{+}$, respectively.

\section{Results and discussion}

\subsection{Cross section for the $\protect\gamma p\rightarrow a_{2}^{+}n$
reaction}

Energy dependence of the cross section calculated above for each of the
models for the fixed parameter $\Lambda _{t}=1$ GeV of the monopole form factor is shown in Fig \ref%
{Fig:total_fix_L}. One can see that the difference between the models is
about an order of magnitude. Fine tuning of the parameter $\Lambda _{t}$
can be performed on the basis of the experimental results.

The existing experimental data \cite{ws75,ye72,mn09,ei69,gt93} for $a_{2}$
photoproduction at low energies are summarized in Table I and Fig. 3.
The original data from \cite{ei69} have been reanalysed in \cite{gt93} to take
into account the actual branching ratio for $a_2\rightarrow 3\pi$ decay channel, since
it was taken to be 100\%. To fix the same problem we scale the result for the cross section
and the error presented in \cite{ye72} by the factor of 1.5. We also skip the result for $E_{\gamma }=19.5$ GeV, which seems are in tension with the shape of the expected theoretical curves.
The MINUIT code of the CERNLIB library was used to perform one-parameter $\chi ^{2}$ fits of the
theoretical curves to the $\sigma _{\gamma p\rightarrow a_{2}^{+}n}$ data.
The free parameters involved and their fitted values are listed in Table II.

\begin{table}[tbph]
\caption{The experimental data for $a_{2}$ photoproduction cross section.
Here the beam energy $E_{\protect\gamma }$ is in the units of GeV, while the
cross section $\protect\sigma _{\protect\gamma p\rightarrow a_{2}^{+}n}$ is
the units of $\protect\mu $b. }%
\begin{tabular}{ccc}
\hline\hline
$E_{\gamma }$ & $\sigma _{\gamma p\rightarrow a_{2}^{+}n}$ & Data source \\
\hline
3.625 (3.25-4.0) & $0.7\pm 0.3$ & \cite{ws75} \\ \hline
4.2 (3.7-4.7) & $1.2\pm 0.45$ & \cite{ye72} \\ \hline
4.8 (4.3-5.25) & $2.6\pm 0.6$ & \cite{gt93,ei69} \\ \hline
5.1 (4.8-5.4) & $0.81\pm 0.25$ & \cite{mn09} \\ \hline
5.15 (4.0-6.3) & $0.3\pm 0.3$ & \cite{ws75} \\ \hline
5.25 (4.7-5.8)& $0.9\pm 0.45$ & \cite{ye72} \\ \hline
7.5 (6.8-8.2)& $0.45\pm 0.45$ & \cite{ye72} \\ \hline
19.5 & $0.29\pm 0.06$ & \cite{gt93} \\ \hline\hline
\end{tabular}%
\end{table}

\begin{table}[tbph]
\caption{The fitted values of the free parameter $\Lambda _{t}$ of the monopole form factor, while the
cutoff $\Lambda _{t}$ is in the units of GeV. }%
\begin{tabular}{ccc}
\hline\hline
type & $\Lambda _{t}$ & $\chi ^{2}/ndf$ \\ \hline
$\pi $ exchange & 0.99$\pm 0.07$ & 2.23 \\ \hline
$\pi $ exchange(Reggeized) & 3.58$\pm 0.66$ & 2.73 \\ \hline
VMD & 0.35$\pm 0.02$ & 2.05 \\ \hline
VMD(Reggeized) & 0.46$\pm 0.04$ & 2.04 \\ \hline\hline
\end{tabular}%
\end{table}

\begin{table}[tbph]
\caption{The fitted values of the free parameter $\Lambda _{d}$ of the dipole form factor, while the
cutoff $\Lambda _{d}$ is in the units of GeV. }%
\begin{tabular}{ccc}
\hline\hline
type & $\Lambda _{d}$ & $\chi ^{2}/ndf$ \\ \hline
$\pi $ exchange & 1.53$\pm 0.11$ & 2.26 \\ \hline
$\pi $ exchange(Reggeized) & 3.79$\pm 0.78$ & 2.91 \\ \hline
VMD & 0.54$\pm 0.03$ & 2.07 \\ \hline
VMD(Reggeized) & 0.69$\pm 0.06$ & 2.05 \\ \hline\hline
\end{tabular}%
\end{table}

\begin{table}[tbph]
\caption{The fitted values of the free parameter $R$ of the exponential form factor, while $R$ is in
the units of GeV$^{-1}$. }%
\begin{tabular}{ccc}
\hline\hline
type & $R$ & $\chi ^{2}/ndf$ \\ \hline
$\pi $ exchange & 0.47$\pm 0.03$ & 2.26 \\ \hline
$\pi $ exchange(Reggeized) & 0.10$\pm 0.02$ & 2.66 \\ \hline
VMD & 1.30$\pm 0.07$ & 2.10 \\ \hline
VMD(Reggeized) & 1.06$\pm 0.09$ & 2.06 \\ \hline\hline
\end{tabular}%
\end{table}

The fitted parameter {$\Lambda _{t}$} satisfys the
expectation with a reasonable $\chi ^{2}/{ndf}$. And yet, it is found that the
cases with Reggeized treatment need {a larger $\Lambda _{t}$}. For
comparison, we also calculate the result with dipole or exponential form
factor. The fitted parameter $\Lambda _{d}$ and $R$ are listed in Table III
and Table IV, respectively. Fig. 3 show that the fitted results with the
three types of form factor. It is found that the difference is small the
result of improvements in three types of form factor. Therefore, in the
following calculation, we only consider the case of monopole form factor.

From Fig. 3(a) one can see that the experimental data (except the point at
$E_{\gamma }=4.8 and 19.5$ GeV)\cite{ws75,ye72,mn09,gt93,ei69} for the total cross
section of the $\gamma p\rightarrow a_{2}^{+}n$ reaction are well reproduced
with {a} small value of $\chi ^{2}{/ndf}$. The shape of the total cross
section via $\pi $ exchange is different from that of the VMD mechanism.

With the above equations and the fitted parameters as listed in Table II,
the relevant physical results are calculated, as shown in Fig. 4-6.

In Fig. 4 we also present the variation of the total cross section of the $%
\gamma p\rightarrow a_{2}^{+}n$ reaction within the typical uncertainties of the $\Lambda _{t}$ values. From Fig. 4 (a) it is seen that the total
cross section via $\pi $ exchange is more sensitive than that of the VMD
mechanism to the values of $\Lambda _{t}$. Moreover, a comparison of the
results from Fig. 4 (a) and Fig. 4 (b) reveals that the total cross
section becomes less sensitive to the $\Lambda _{t}$ values when the
Reggeized treatment is added to the process of $\gamma p\rightarrow
a_{2}^{+}n$.

In Fig. 5, we show the differential cross section of $\gamma p\rightarrow
a_{2}^{+}n$ as a function of $-t$. It is obvious that there is a
significant peak structure in the region of low $-t$, which increases
rapidly near the threshold and then decreases slowly {with increasing} $-t$.
However, it is seen, that the shapes of the differential cross section $%
d\sigma /dt$ with the Reggeized treatment are much different from that
without the Reggeized treatment at higher $-t$. The Reggeized treatment
can lead to that the differential cross section $d\sigma /dt$ decreases
rapidly with increasing $-t$, especially at higher energies.

Figure 6 presents the differential cross section for the $\gamma
p\rightarrow a_{2}^{+}n$ reaction with or without the Reggeized treatment
at different energies. It is seen that the differential cross section with
the Reggeized treatment is very sensitive to the angle $\theta $ and makes a considerable contribution at forward angles.

\begin{figure}[tbp]
\begin{center}
\includegraphics[scale=0.4]{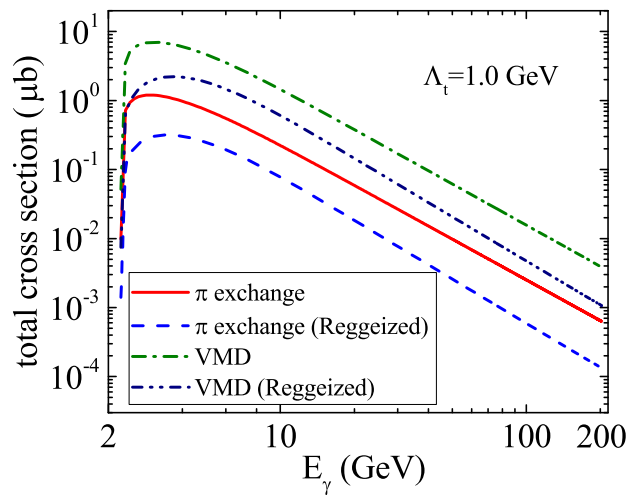}
\end{center}
\caption{(Color online) Total cross section for {the} $\protect\gamma %
p\rightarrow a_{2}^{+}n$ reaction for the fixed parameter $\Lambda _{t}=1$
GeV.}
\label{Fig:total_fix_L}
\end{figure}

\begin{figure}[tbp]
\begin{center}
\includegraphics[scale=0.4]{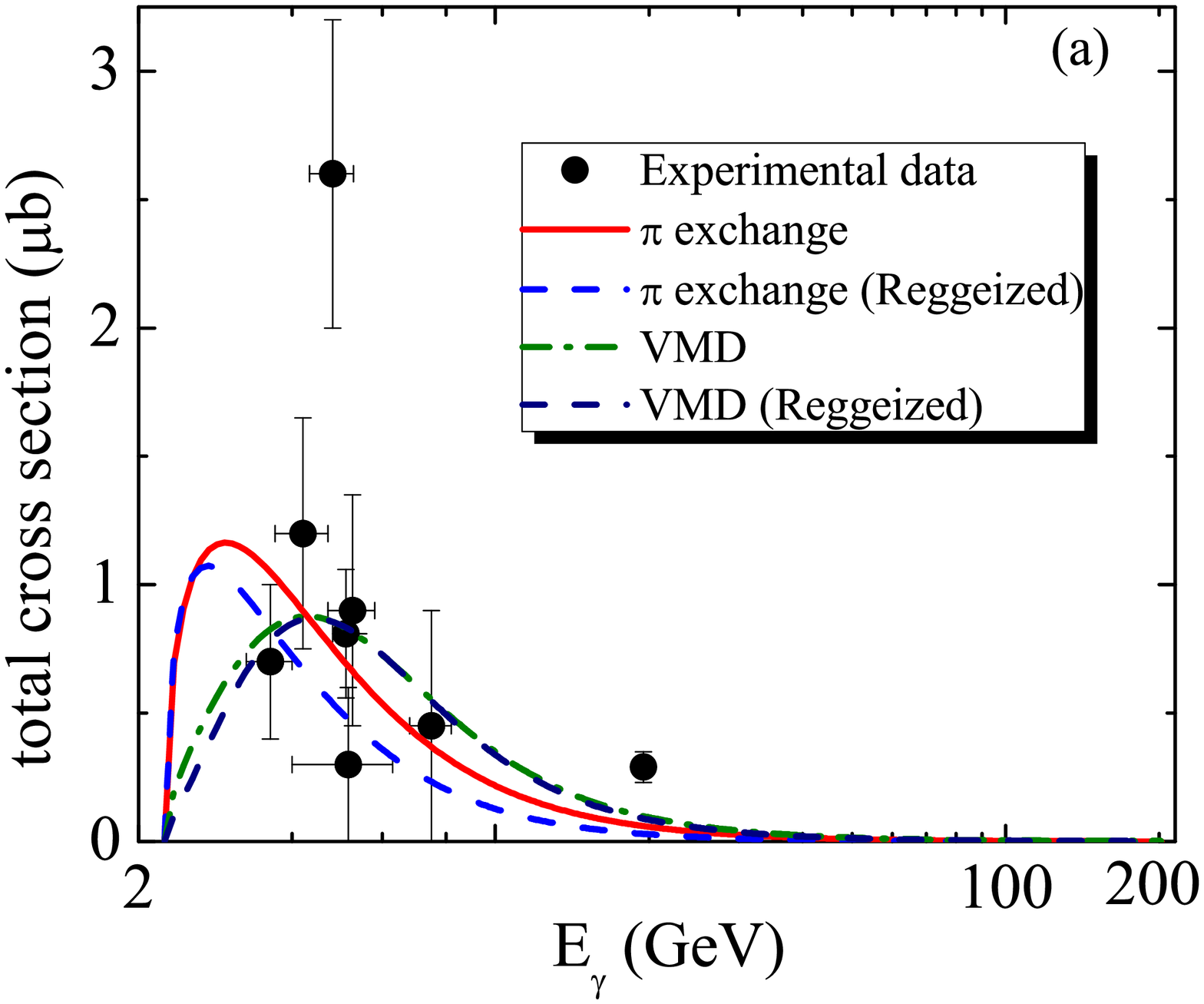} \includegraphics[scale=0.4]{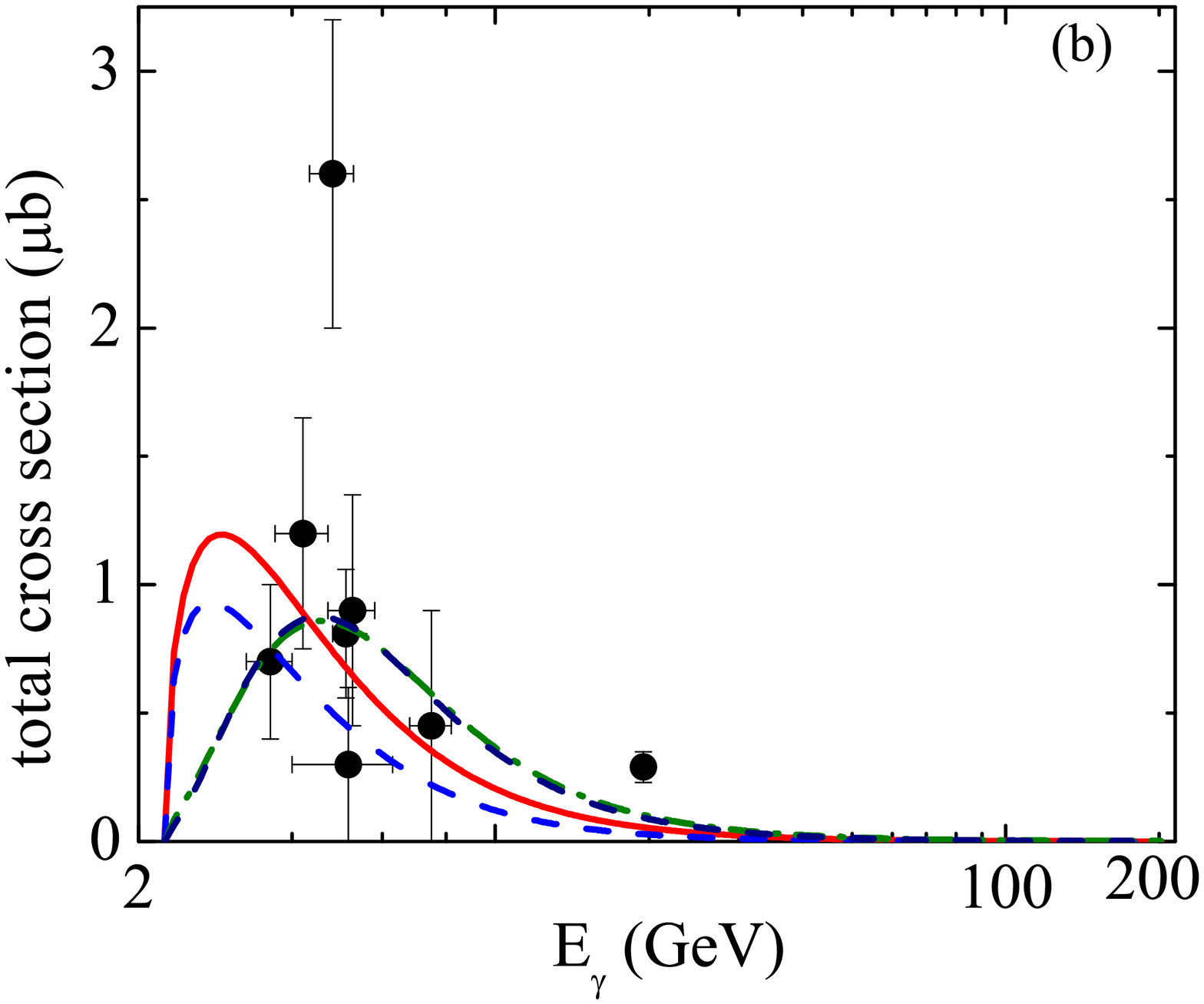} %
\includegraphics[scale=0.4]{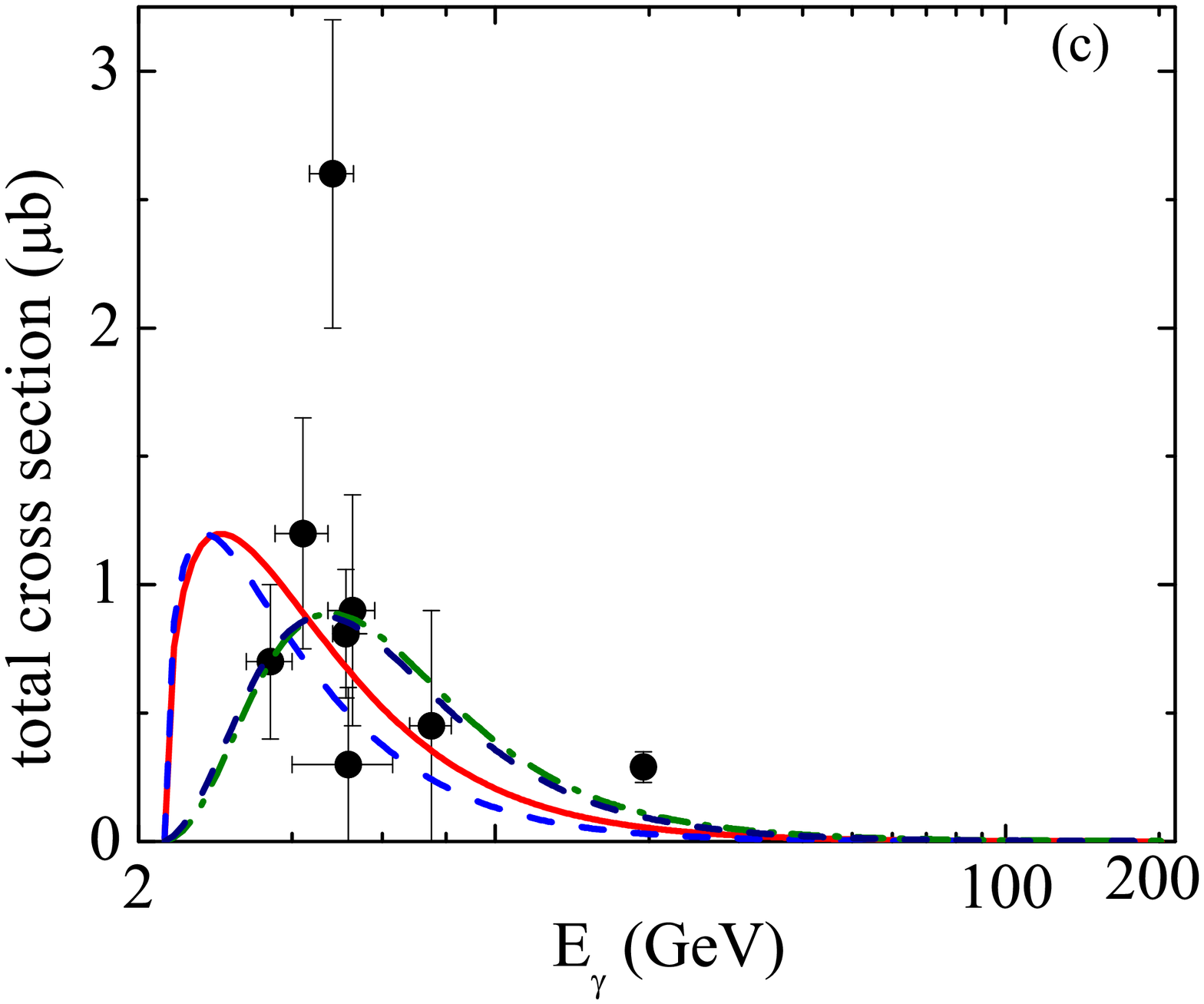}
\end{center}
\caption{(Color online) Total cross section for {the} $\protect\gamma %
p\rightarrow a_{2}^{+}n$ reaction. The datas are from \protect\cite%
{ws75,ye72,mn09,gt93,ei69}. Here, (a) is the result with monopole form factor,
while (b) and (c) are the results related to the dipole and exponential form
factors correspondently.}
\label{Fig:total}
\end{figure}

\begin{figure}[tbp]
\begin{center}
\includegraphics[scale=0.4]{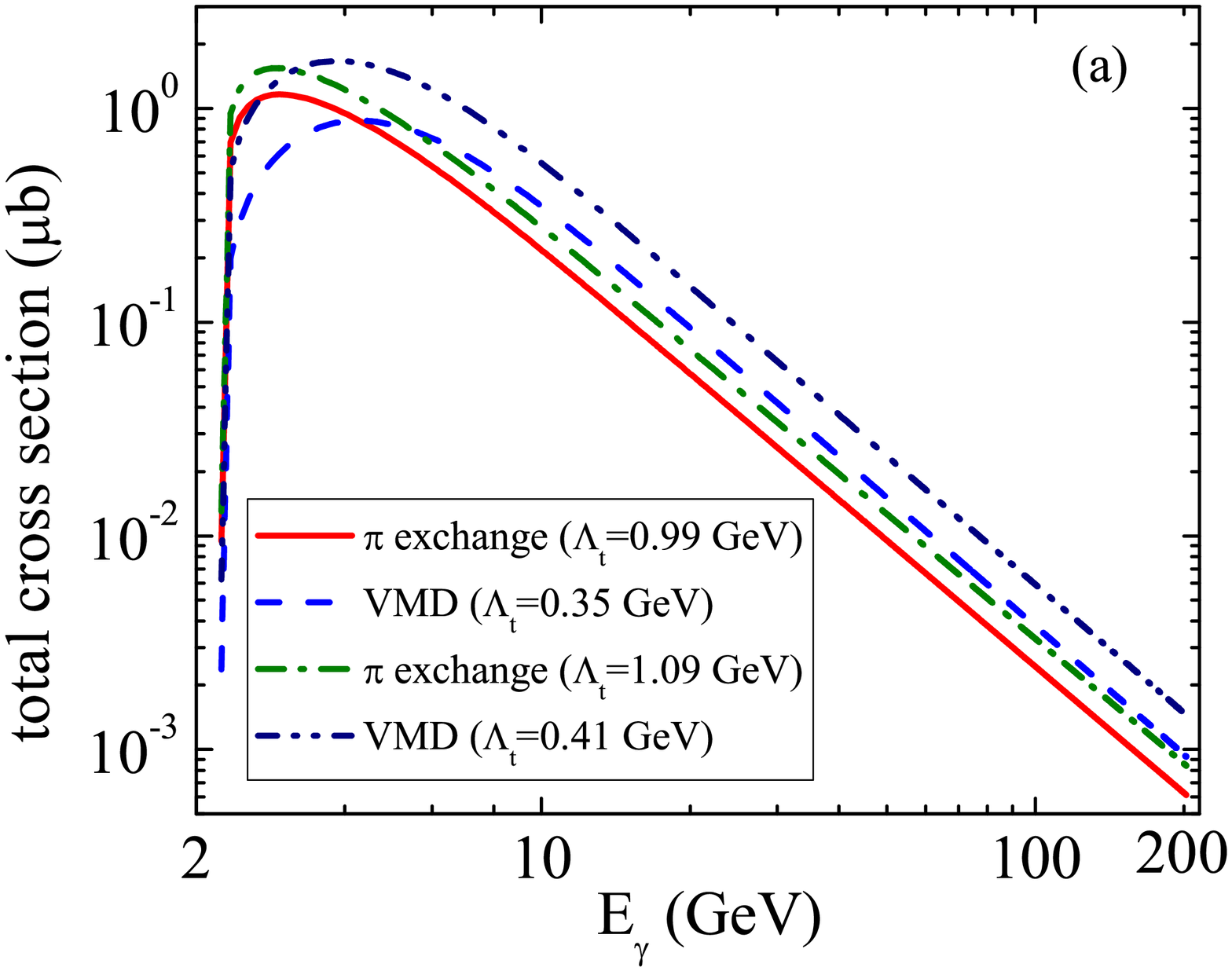} \includegraphics[scale=0.4]{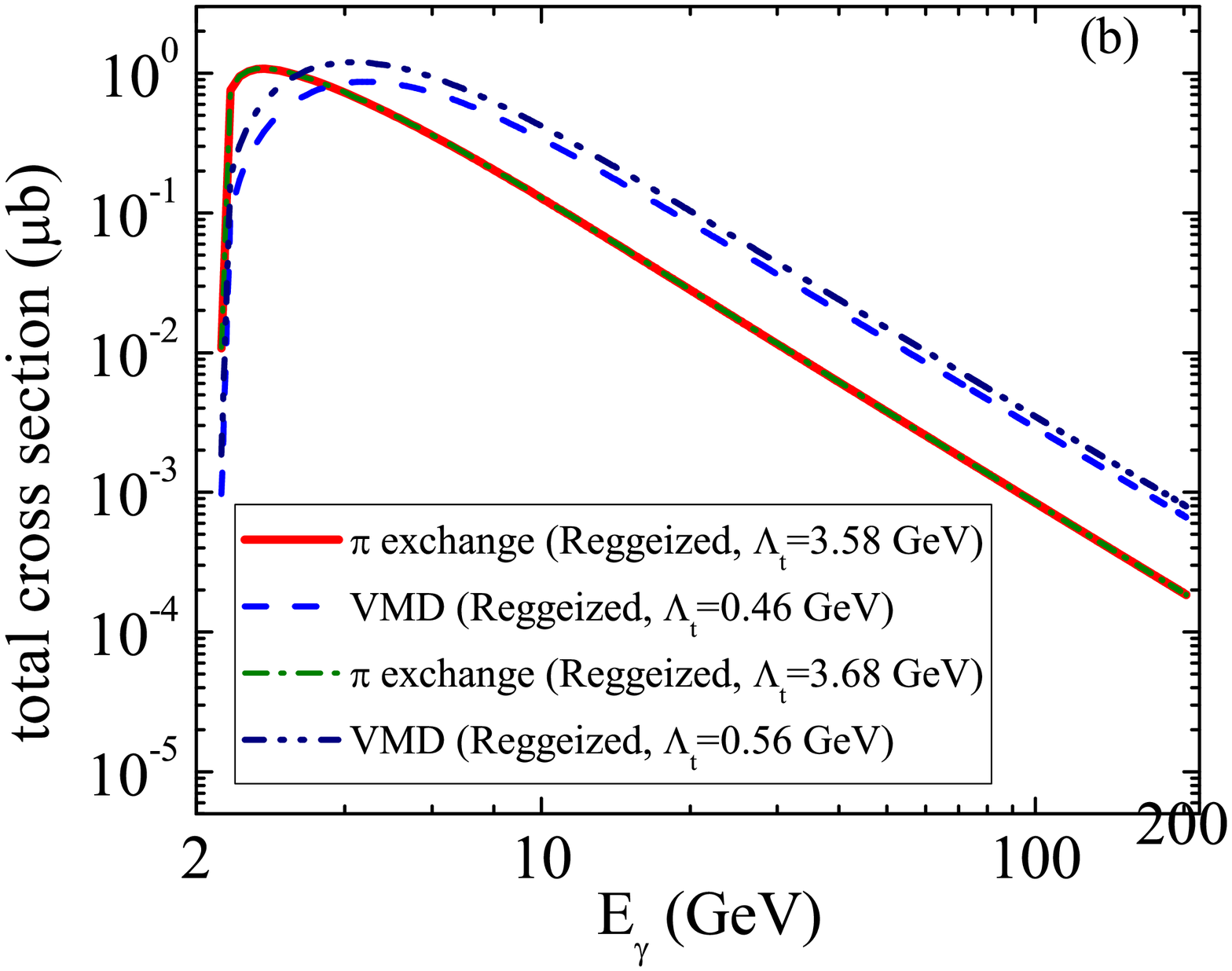}
\end{center}
\caption{(Color online) (a): {The total cross section of the $\protect\gamma %
p\rightarrow a_{2}^{+}n$ reaction for} the different values of cutoff
parameter $\Lambda _{t}$. (b){\ same as in} (a), but for the case of {the}
Regge trajectory exchange.}
\label{Fig:totalv}
\end{figure}

\begin{figure*}[t]
\begin{center}
\includegraphics[scale=0.8]{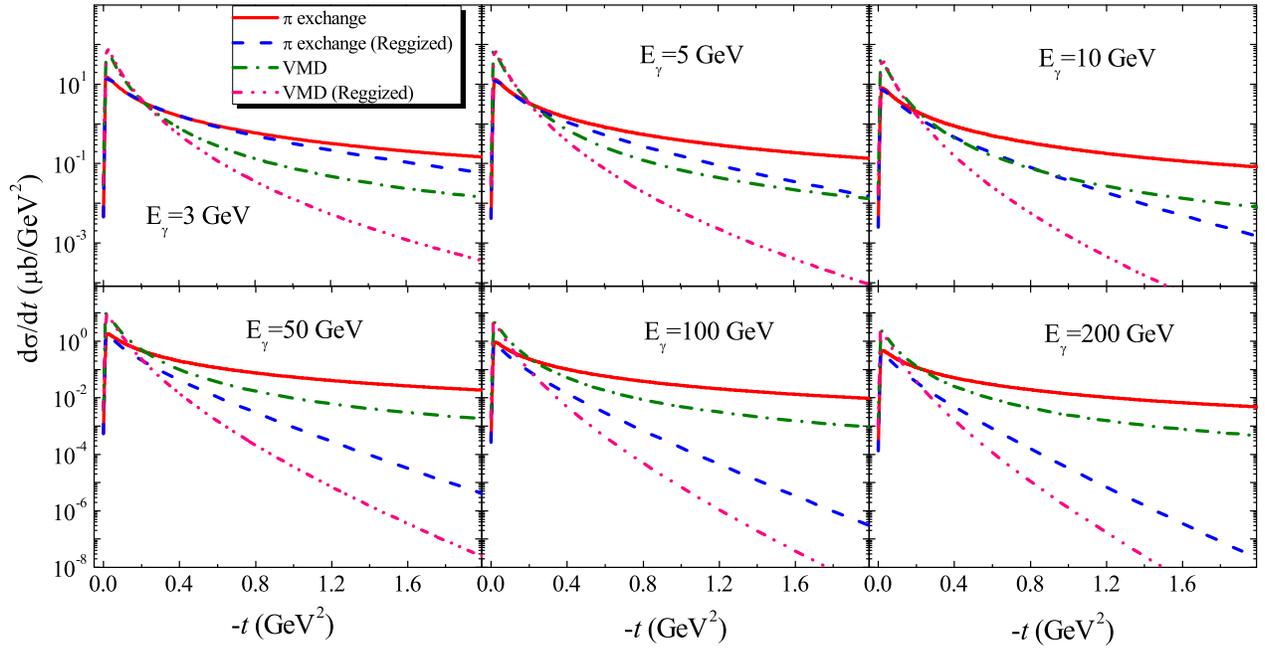}
\end{center}
\caption{(Color online) The differential cross section of $\protect\gamma %
p\rightarrow a_{2}^{+}n$ {as a function of} $-t$ at $E_{\protect\gamma %
}=3-200$ GeV.}
\end{figure*}

\begin{figure*}[t]
\begin{center}
\includegraphics[scale=0.8]{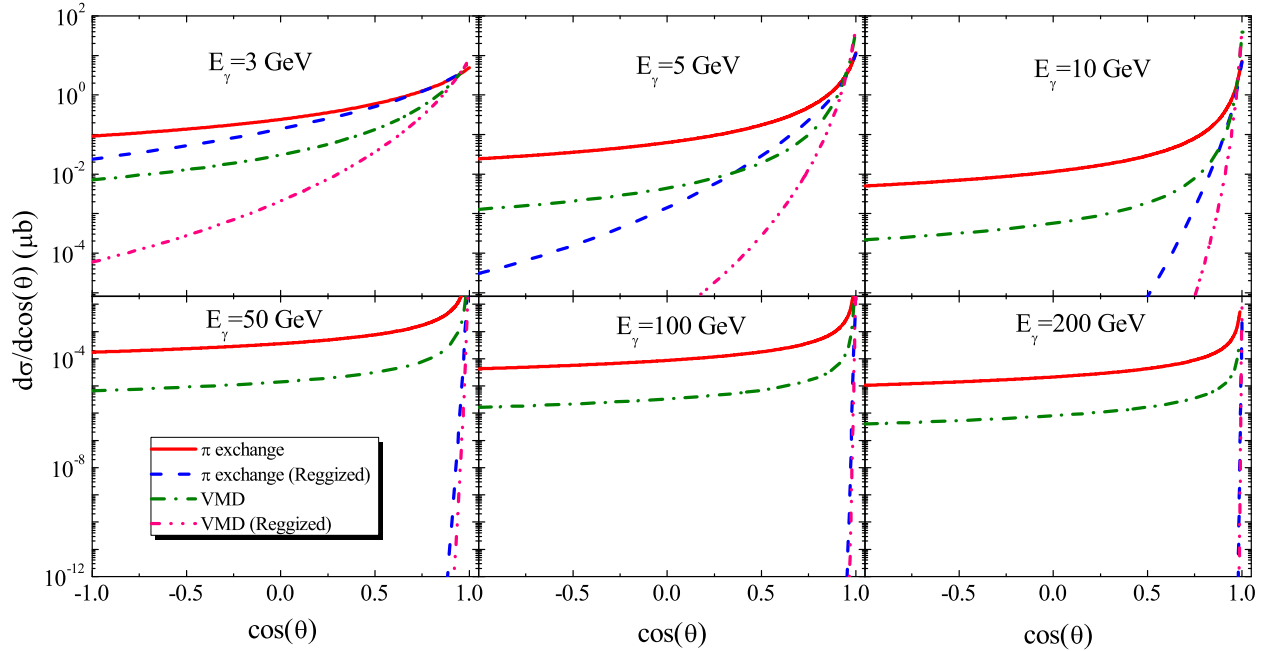}
\end{center}
\caption{(Color online) The differential cross section $d\protect\sigma %
/d\cos \protect\theta $ for the $a_{2}(1320)$ photoproduction from the
protron as a function of $\cos \protect\theta $ at $E_{\protect\gamma %
}=3-200 $ GeV.}
\end{figure*}

\begin{figure*}[t]
\begin{center}
\includegraphics[scale=0.8]{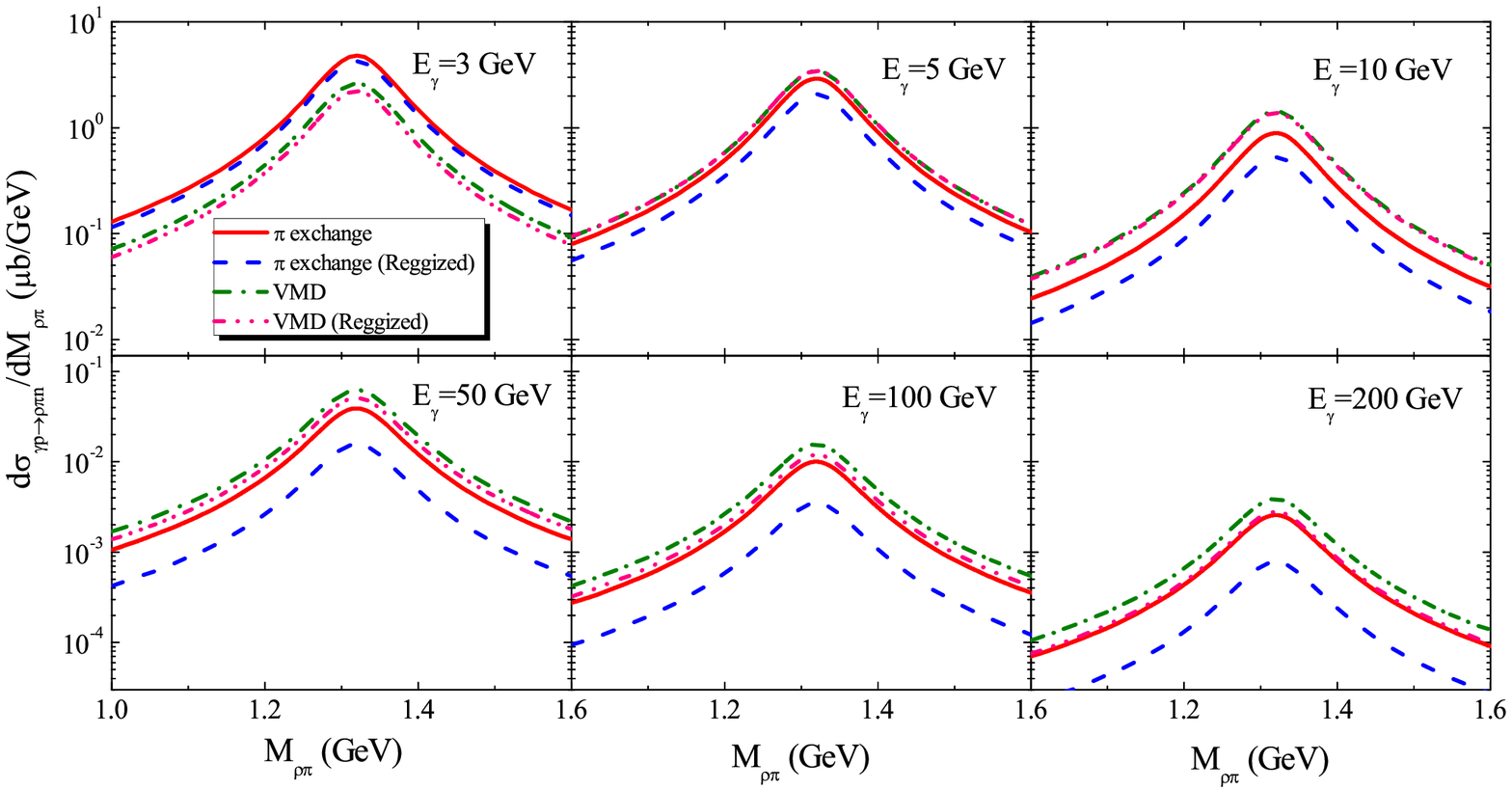}
\end{center}
\caption{(Color online) Differential cross section $d\protect\sigma _{%
\protect\gamma p\rightarrow a_{2}^{+}n\rightarrow \protect\rho \protect\pi %
n}/dM_{\protect\rho \protect\pi }$ as a function of $M_{\protect\rho \protect%
\pi }$ at $E_{\protect\gamma }=3-200$ GeV.}
\end{figure*}

\subsection{Dalitz process $\protect\gamma p\rightarrow \protect\rho \protect%
\pi n$}

Considering the $a_{2}$ are usually detected in experiment via the $\rho
\pi $ invariant mass , it would be useful to give the theoretical
predictions of the differential cross section $d\sigma _{\gamma p\rightarrow
a_{2}^{+}n\rightarrow \rho \pi n}/dM_{\rho \pi }$ as a function of the
beam energy $E_{\gamma }$, which could be tested by further experiment.
Since the full decay width of the $a_{2}(1320)$ is small enough in
comparison to its mass, the invariant mass distribution for the Dalitz
process $\gamma p\rightarrow \rho \pi n$ can be defined{\ with{\ the}
two-body process} \cite{sn15}%
\begin{equation*}
\frac{d\sigma _{\gamma p\rightarrow \rho \pi n}}{dM_{\rho \pi }}\approx
\frac{2m_{a_{2}}M_{\rho \pi }}{\pi }\frac{\sigma _{\gamma p\rightarrow
a_{2}^{+}n}\Gamma _{a_{2}\rightarrow \rho \pi }}{(M_{\rho \pi
}^{2}-m_{a_{2}}^{2})^{2}+m_{a_{2}}^{2}\Gamma _{a_{2}}^{2}},
\end{equation*}%
where the full width $\Gamma _{a_{2}}=107$ MeV and{\ the} partial width $%
\Gamma _{a_{2}\rightarrow \rho \pi }=75$ MeV are taken \cite{pdg}.

With the above equations and the fitted parameters as listed in Table II,
the invariant-mass distribution $d\sigma _{\gamma p\rightarrow
a_{2}^{+}n\rightarrow \rho \pi n}/dM_{\rho \pi }$ for $E_{\gamma }=3-200$
GeV is calculated, as shown in Fig. 7. It is seen that there exists an
obvious peak at $M_{\rho \pi }\approx 1.32$ GeV.

\section{Possibility of the experimental test at COMPASS{\ }}

The COMPASS experiment \cite{Abbon:2007pq} is situated at the M2 beam line
of the CERN Super Proton Synchrotron. Since 2002 it {has} obtained
experimental data for positive muons scattering of 160~$GeV/c$ (2002-2010)
or 200~$GeV/c$ momentum (2011) off solid $^{6}$LiD (2002-2004) or NH$_{3}$
polarized targets (2006-2011). Particle tracking and identification is
performed in a two-stage spectrometer covering a wide kinematical range. The
trigger system comprises hodoscope counters and hadron calorimeters.

According to the presented calculations of the $a_{2}$ production cross
section and previously published COMPASS results for exclusive
photoproduction of $\rho ^{0}$ \cite{rho0} and $J/\psi $ \cite{Jpsi} we can
conclude that thousands of $a_{2}^{\pm }$ mesons could be produced per year
of data taking via the exclusive {charge-exchange reactions} $\gamma ^{\ast
}p\rightarrow a_{2}^{+}n$ and $\gamma ^{\ast }n\rightarrow a_{2}^{-}p$ (but
the recoil nucleon cannot be detected). The energy of a virtual photon
covers the range from about 20 GeV and up to 180 GeV. The obtained data can
be used to clarify the mechanism of the $a_{2}$ production and the role of
the Reggeized treatment at high energies. Nevertheless{, most of the} $%
a_{2}^{\pm }$ mesons at such energies are produced non exclusively via the
pomeron exchange mechanism. Such events could produce strong background {%
under} poor exclusivity control. Additional systematics could come from the
process $\gamma ^{\ast }p\rightarrow a_{2}^{-}\Delta ^{++}$ (the cross
section of this reaction is of the same order of magnitude \cite{DeltaPP})
because the COMPASS setup is not able to reconstruct a decay of low-energy
$\Delta ^{++}$ in a regular way. Nuclear effects in $a_{2}$
photoproduction off the lithium-6, deuterium and nitrogen nuclei should
also be taken into account in an appropriate way.

The forthcoming upgrade of the COMPASS setup related to the planned
data taking within the framework of the GPD program \cite{COMPASS_proposal}
could provide better conditions for experimental study of the reaction $%
\gamma ^{\ast }p\rightarrow a_{2}^{+}n$ and partially eliminate problems
mentioned above. The new 2.5 m long liquid hydrogen target surrounded by a 4
m long recoil proton detector will be used. Absence of the neutrons in the
target will remove one exclusive production channel for $a_{2}^{-}$. The
recoil proton detector serves the double purpose: to reconstruct and
identify recoil protons via time-of-flight and energy loss measurements.
Since the reaction $\gamma ^{\ast }p\rightarrow a_{2}^{+}n$ does not have
a recoil proton in the final state, any activity in the recoil proton
detector could be used as the veto in the offline analysis. In addition,
the recoil proton detector will be able to detect the $\Delta
^{++}\rightarrow p\pi ^{+}$ decay. Significant impact on the exclusivity
control efficiency will be made by the planned upgrade of the
electromagnetic calorimetry system.

\section{Summary}

Within the framework of the effective Lagrangian approach and Regge model,
the $a_{2}(1320)$ photoproduction from the proton is investigated.

The obtained numerical results indicate the following:

\begin{itemize}
\item[(I)] The total cross section $\gamma p\rightarrow a_{2}^{+}n$ related
to the experimental data \cite{ws75,ye72,mn09,ei69,gt93} is well reproduced
with reasonable value of $\chi ^{2}{/ndf}$. Although the monopole form factor
was used in the calculations, the dipole and exponential form factors were also tested. It is found that the total cross
section becomes less sensitive to the $\Lambda _{t}$ values when the
Reggeized treatment is added.

\item[(II)] The shapes of the differential cross section $d\sigma /dt$
with the Reggeized treatment are much different from that without the
Reggeized treatment at higher $-t$. The Reggeized treatment can lead to that the differential cross section $d\sigma /dt$ decreases rapidly with the
increasing $-t$, especially at higher energies.

\item[(III)] The differential cross section $d\sigma /d\cos \theta $ with
the Reggeized treatment is very sensitive to the angle $\theta $ and makes
a considerable contribution at forward angles.

\item[(IV)] The invariant mass distribution for the Dalitz process $\gamma
p\rightarrow \rho \pi n$ shows an obvious peak at $M_{\rho \pi }\approx
1.32$ GeV, which can be checked by further experiment.

\item[(V)] For comparison, it is found that the cross section of the $%
\gamma n\rightarrow a_{2}^{-}p$ process is almost the same as that of the $%
\gamma p\rightarrow a_{2}^{+}n$ reaction. Thus, the above theoretical
results are valid to the $\gamma n\rightarrow a_{2}^{-}p$ channel.
\end{itemize}

To sum up, we suggest testing our prediction for the cross section of
the $\gamma p\rightarrow a_{2}^{+}n$ process at the COMPASS facility at
CERN. Such a test could provide important information for clarifying the
production mechanism of the $a_{2}(1320)$ and the role of the Reggeized
treatment at high energies. Nevertheless, the precise measurements near
the threshold, where the difference between the predictions of the
production models is maximal, are also important.

\section{Acknowledgments}

The author X. Y. Wang is grateful Helmut Haberzettl and Jun He for useful
discussions about Regge theory.

\end{document}